\documentclass[twocolumn]{aastex62}

\newcommand{\beq}{\begin{equation}}
\newcommand{\eeq}{\end{equation}}
\newcommand{\erf}{{\rm erf}}
\newcommand{\erfc}{{\rm erfc}}
\newcommand{\Msun}{M_{\odot}}
\newcommand{\Lsun}{L_{\odot}}
\newcommand{\Duty}{\mathcal{D}}
\newcommand{\fedd}{f_{\rm Edd}}
\newcommand{\tedd}{t_{\rm E}}
\newcommand{\Ledd}{L_{\rm E}}
\newcommand{\Lacc}{L_{\rm acc}}

\newcommand{\Mdotacc}{\dot{M}_{\rm acc}}
\newcommand{\ndcbh}{n_{\rm DCBH}}

\newcommand{\Jcrit}{J_{\rm crit}}
\newcommand{\melec}{m_{\rm e}}
\newcommand{\mpro}{m_{\rm p}}

\graphicspath{{./}{figures/}}

\shorttitle{SMBH Mass Function}
\shortauthors{Basu \& Das}
\begin{document}

%\title{THE MASS FUNCTION OF DIRECT COLLAPSE BLACK HOLES}
\title{THE MASS FUNCTION OF SUPERMASSIVE BLACK HOLES IN THE DIRECT COLLAPSE SCENARIO}

%% you desire. In v6.2 you can include a footnote in the title.

\author[0000-0003-0855-350X]{Shantanu Basu}
\affiliation{Department of Physics and Astronomy, University of Western Ontario, London, Ontario N6A 3K7, Canada}

\author{Arpan Das}
\affiliation{Department of Physics and Astronomy, University of Western Ontario, London, Ontario N6A 3K7, Canada}

\correspondingauthor{Shantanu Basu}
\email{basu@uwo.ca}
\begin{abstract}
	One of the ideas to explain the existence of supermassive black holes (SMBH) that are in place by $z \sim 7$ is that there was an earlier phase of very rapid accretion onto direct collapse black holes (DCBH) that started their lives with masses $\sim 10^{4-5}\, \Msun$. Working in this scenario, we show that the mass function of SMBH after such a limited time period with growing formation rate paired with super-Eddington accretion can be described as a broken power-law with two characteristic features. There is a power-law at intermediate masses whose index is the dimensionless ratio $\alpha \equiv \lambda/\gamma$, where $\lambda$ is the growth rate of the number of DCBH during their formation era, and $\gamma$ is the growth rate of DCBH masses by super-Eddington accretion during the DCBH growth era. A second feature is a break in the power law profile at high masses, above which the mass function declines rapidly. The location of the break is related to the dimensionless number $\beta = \gamma\,T$, where $T$ is the duration of the period of DCBH growth. If the SMBH continue to grow at later times at an Eddington-limited accretion rate, then the observed quasar luminosity function can be directly related to the tapered power-law function derived in this paper.
%These features can also appear in the luminosity function since Eddington-limited accretion leads to a direct relation between mass and luminosity. The truncated power-law function dervied here can therefore be used to model the observed quasar luminosity function observed at lower redshifts. 

\end{abstract}



\section{Introduction} \label{sec:intro}

A key challenge to the theory of the formation of supermassive black holes (SMBH)
in the early universe is the observation of very massive ($M \approx 10^9 \Msun$)
and luminous ($L \gtrsim 10^{13}\,\Lsun$) quasars already in place by $z \sim 7$, when the 
universe is just $\sim 800$ Myr old \citep[e.g.,][; see also the
review by \citet{woo18}]{fan06,mor11,wu15,ban18}. Objects that accumulate 
at least a billion $\Msun$
in less than a billion years after the Big Bang put a strain on the normal ideas of 
Eddington-limited growth of black hole seeds that originate from Population III stellar
remnants. Starting from a seed mass $M_0$, Eddington-limited growth leads to a mass
$M(t) = M_0\exp[\epsilon^{-1}(1-\epsilon)t/\tedd]$,
where $\tedd \approx 450$ Myr and $\epsilon$ ($\approx 0.1$)
is a radiative efficiency factor. Population III stars are thought 
to have masses $\lesssim 40 \Msun$ \citep{hos11}, and their remnants would be 
less massive, so that
there is apparently not enough time available to reach $M \sim 10^9\Msun$. 
These constraints show that a combination of both more massive initial seeds and a 
super-Eddington growth rate may be necessary to account for the observed SMBH at 
$z \sim 7$.

One promising pathway is that of direct collapse black holes (DCBH) \citep{bro03}. 
The idea is that Lyman-Werner (LW) photons (having energies 11.2 to 13.6 eV) 
from the first Population III
stars can propagate far from their sources and dissociate H$_2$ in other primordial
gas clouds. Without H$_2$ cooling these gas clouds equilibrate to temperatures
$T \sim 8000$ K set by atomic cooling, which means that the Jeans mass is 
$\sim 10^5\Msun$ at a number density $n = 10^4$ cm$^{-3}$, as opposed to $\sim 10^3\Msun$ in a normal Population III star formation
environment or $\sim 1 \Msun$ in present day star formation. Due to their large
mass these collapsing cloud fragments may be able to collapse directly into black
holes, after a brief period as a supermassive star \citep{bro03} or quasi-star 
\citep{beg06,beg08}, if the infalling matter can overcome the angular
momentum barrier and disruptive effects of radiative feedback. An interesting joint solution 
to these barriers is proposed by \citet{sak16} based on 
the episodic accretion scenario of \citet{vor13} that is powered by gravitational
instability in a circumstellar disk. In this model the episodic
accretion results in a lower surface temperature of a supermassive star, thereby
also reducing the effect of radiative feedback that can limit mass accumulation in the 
case of normal Population III star formation \citep{hos11}. The DCBH model has been 
extensively developed
in the context of galaxy formation models, resulting in a scenario where the formation
of atomic cooling halos is seeded by the first stars, and the subsequent DCBH
produce LW radiation that triggers the formation of other atomic cooling halos and DCBH 
in a kind of chain reaction process \citep{yue14}. A rapid
period of growth of atomic cooling halos and therefore DCBH formation ensues, 
with the growth rate at any time related to the instantaneous number of DCBH. The 
rapidly growing phase of DCBH creation is also a period of possible rapid mass growth 
through super-Eddington accretion \citep{ina16,pac17}. The whole process comes to a 
rapid halt however, when the gas in the atomic cooling halos is photoevaporated 
by the ambient radiation field. According to \citet{yue14} the DCBH era lasts from
$z \approx 20$ to $z \approx 13$, or a time period $T \approx 150$ Myr, after which DCBH
formation is completely suppressed. Here we adopt the picture emerging from their
semi-analytic model, however we note that numerical simulations \citep[e.g.,][]{aga12,cho16,hab16} have not reached a consensus on the DCBH formation rate or the termination time of their formation.
%delete "simple" below?

In this Letter, we seek a simple model of the growth of DCBH in the early universe that 
captures just the essential features of the scenario described above, in order to 
reach an analytic understanding of the mass and luminosity functions of observable 
quasars that form through the DCBH scenario. 
%sakurai

%Observational motivation\\
%- DCBH idea\\
%- super-Eddington growth\\
%- chain reaction of DCBH creation\\
%- termination of creation and rapid accretion\\
%Follow scenario described in \citet{yue14}.

%The main challenge for the stellar seed BH mechanism is the ~ 7 orders of magnitude
%growth in mass that is required in 700 - 800 Myr. However, if one is to assume a larger
%seed mass, say Mseed ~ 104-5 M?, then the growth in mass required is only ~ 4 - 5 orders
%of magnitude.

%Another idea has been proposed in order to create massive seed BHs out of pristine gas
%with either no star formation in the previous stages, or a very short lived supermassive
%(or quasistar) star phase (Oh & Haiman, 2002).

\section{Background} \label{sec:growth}

%The DCBH formation scenario emerging from numerical simulations 
%\citep[e.g.,][]{sha10,fer14,hab16} is 
%of an early limited time period with a growing rate of DCBH formation, followed
%by a rapid termination of DCBH formation at a certain time. 
\citet{yue14}
estimate that the rapid formation of DCBH occurs between $z \approx 20$
and $z \approx 13$, after which it ends abruptly. The semi-analytic model of 
\citet{dij14} is broadly consistent with this and shows that the number density of DCBH
$\ndcbh$ grows rapidly during a similar interval.

\begin{figure}[h]
	\plotone{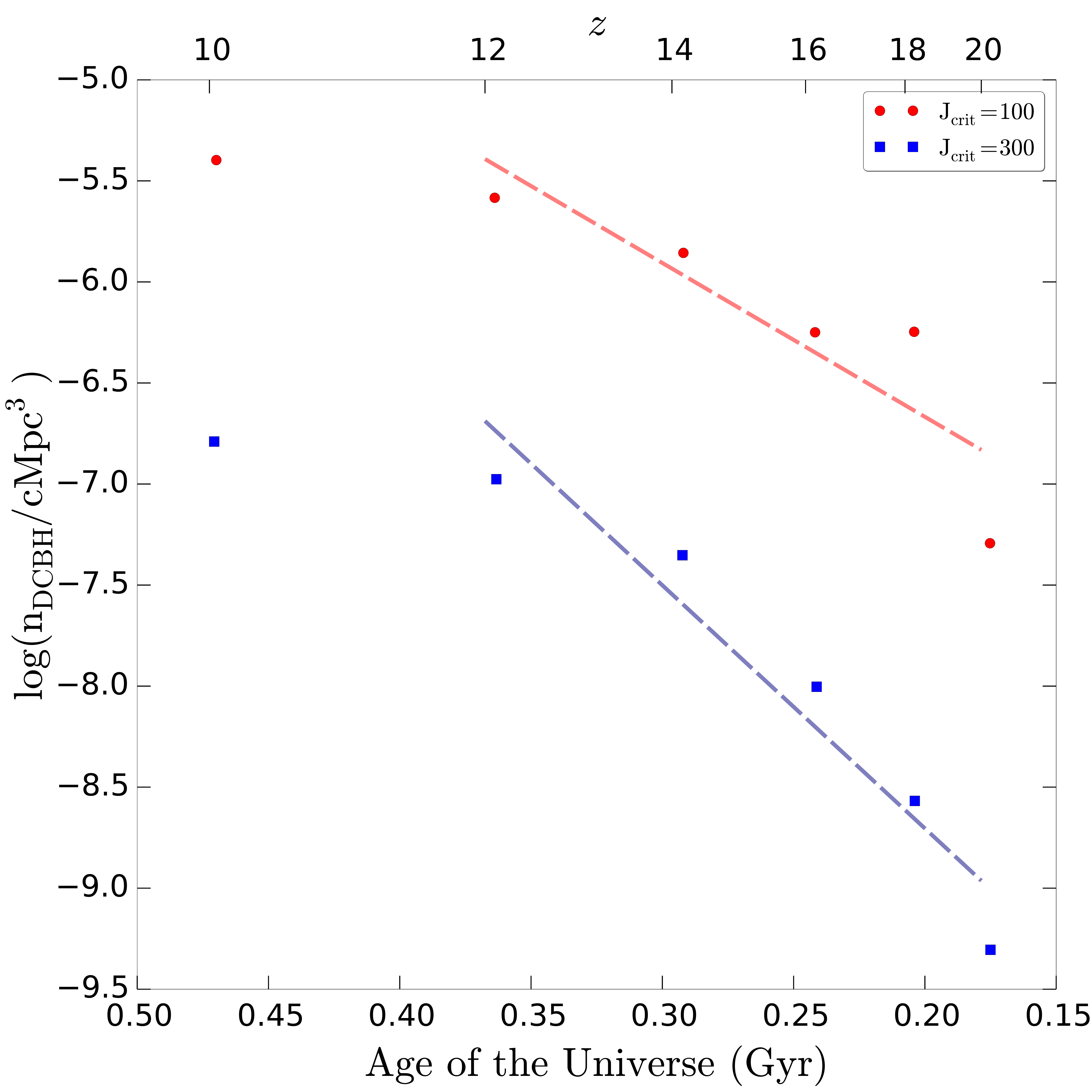}
	\caption{The growth of the number density of DCBH $\ndcbh$. The data points correspond to $\ndcbh$ (in cMpc$^{-3}$) at redshift values $z=20.3, 18.2, 16.2, 14.1, 12.1, 10.0$ corresponding to cosmic times $0.18, 0.20, 0.24, 0.29, 0.36, 0.47$ Gyr after the Big Bang, and are taken from \citet{dij14}.}
	\label{fnum}
\end{figure}

Figure \ref{fnum} shows the results of \citet{dij14} 
plotted against time after the Big Bang, for two possible values of the 
critical flux of LW photons $\Jcrit$ (written in units of $10^{-21}$ erg s$^{-1}$ cm$^{-2}$ Hz$^{-1}$ sr$^{-1}$) that is needed to create the atomic cooling
halos that are DCBH progenitors. The calculation of $\Jcrit$ \citep[e.g.,][]{sha10}
includes the additional effect of near-infrared radiation with energy above 0.76 eV
that can also inhibit the formation of H$_2$ by dissociating the H$^{-}$ ion that
is an intermediary in the H$_2$ formation chain.
Although significant uncertainties exist in the actual number of DCBH created, 
due to uncertainties in the value of $\Jcrit$
that depends on the source density and spectra, escape fraction from the host halos, etc.,
the slope of growth, $\lambda(t) \equiv d \ln \ndcbh/dt$, is similar in all 
their modeled cases. For the two cases shown here, we find an average
value $\lambda$ measured between data points from $z=20$ to $z=12$, which 
corresponds roughly to the period of rapid DCBH formation. 
For their canonical model $\Jcrit=300$ the best fit line yields 
$\lambda=27.7$ Gyr$^{-1}$, and 
for $\Jcrit=100$ the best fit is $\lambda =17.7$ Gyr$^{-1}$.
Here we use the canonical model and 
adopt $\lambda = 28.0$ Gyr$^{-1}$, slightly steeper than the best fit.
This in the interest of rounding off and also because there is evidence that the
DCBH growth era ended just prior to this data point, at $z \approx 13$ \citep{yue14}, 
which would tend to drive the slope to a slightly greater value.

Each DCBH can be modeled as growing in mass at an exponential rate, but the 
starting times of the accretion process will be spread throughout the DCBH
formation era. However, the super-Eddington growth will cease for all 
DCBH at about the same time, so that there will be a distribution of accretion
times among the population of DCBH. This is a key part of our model as
developed in Section \ref{sec:massfcn}.

The growth of an individual DCBH is thought to proceed by default at an 
Eddington-limited rate, but periods of super-Eddington growth are also 
possible \citep{pac17}. 
%The Eddington-limited growth is an example of 
%an exponential growth of mass. 
The Eddington luminosity is 
\beq
\Ledd = \frac{4 \pi c G \mpro M}{\sigma_T}.
\label{Ledd}
\eeq
where $M$ is the black hole mass, $\mpro$ is the proton mass, and 
$\sigma_T = (8\pi/3)(e^2/\melec c^2)^2$ is the Thomson cross section in which
$\melec$ is the electron mass.
At this luminosity the radiation pressure can balance the gravitational 
pressure. The accretion of mass to very small radii comparable to the Schwarzschild
radius will release a significant portion of the rest mass energy, hence
the luminosity is normally estimated as $\Lacc = \epsilon \Mdotacc c^2$, where
$\Mdotacc$ is the mass accretion rate and $\epsilon$ is the radiative efficiency, 
typically set to 0.1. Since the accretor will gain 
rest mass at the rate $\dot{M} = (1-\epsilon)\Mdotacc$, we equate $\Lacc$
with $\Ledd$ to find that   
%and the mass $M$ and radius $R$ of the accretor by
%$\Lacc = \epsilon G M \dot{M}/R$ where $\epsilon$ is the radiative efficiency, 
%typically set to 0.1. Equating $\Lacc$ to $\Ledd$ and using the 
%Schwarzschild radius $R = 2 G M/c^2$ we find that
\beq
\frac{dM}{dt} = \gamma_0 \, M \Rightarrow M(t) = M_0 \exp(\gamma_0 t),
\eeq
where $\gamma_0 = (1-\epsilon)/(\epsilon \tedd)$ and $\tedd = 2e^4/(3G\mpro\melec^2c^3) = 450$ Myr is the Eddington time. 
Here we follow \citet{pac17} in accounting for the idea 
that accretion (especially of the super-Eddington variety) may be episodic, by 
identifying the duty cycle $\Duty$ as the  
fraction of time spent accreting, and the Eddington ratio $\fedd$ that is $=1$ for Eddington-limited
accretion but can be $<1$ for sub-Eddington accretion and $>1$ for super-Eddington 
accretion.  
We use a generalized accretion rate
$\gamma = \chi \gamma_0$ where
$\chi = \Duty \fedd$ is a correction factor to account for the fact that 
the accretion rate could be super-Eddington for
some periods of time. 
The quantitatively relevant parameter is $\chi = \Duty\fedd$. \citet{pac17}
find that objects with $M \gtrsim 10^5 \Msun$ can have high efficiency accretion, 
$0.5\leq \Duty \leq 1$ and $1\leq \fedd \leq 100$, but objects with $M\lesssim 10^5 M_\odot$
have low efficiency accretion, $0\leq \Duty\leq 0.5$ and $0\leq \fedd \leq 1$.
The simplest assumption is that $\chi=1$ for Eddington-limited growth, but our 
model allows for the putative super-Eddington growth in the DCBH formation era.

\section{Mass Function}
\label{sec:massfcn}

We assume that the distribution of initial black hole masses is lognormal, i.e., the differential number density per logarithmic mass bin is distributed normally:
\beq 
\frac{dn}{d\ln M_0} = \frac{1}{\sqrt{2\pi}\sigma_0}\exp{\left(-\frac{(\ln M_0 - \mu_0)^2}{2\sigma_0^2}\right)}.
\eeq
Here $\mu_0$ and $\sigma_0$ are the mean and standard deviation of the distribution of $\ln M_0$, respectively. A lognormal distribution for the birth mass function of DCBH
seeds is consistent with the results of \citet{fer14} for intermediate masses
($4.75 < \log (M/\Msun) < 6.25$), and we fit those results with 
$\mu_0 = 11.7$ (corresponding to a peak at $\log M/\Msun = 5.1$) and $\sigma_0=1.0$.
%need a bit more discussion above

Since the growth law implies that 
\beq  
\ln M(t) = \ln M_0 + \gamma\,t,
\eeq
we can write the mass function at a later time as 
\beq  
\frac{dn}{d\ln M} = \frac{1}{\sqrt{2\pi}\sigma_0}\exp{\left(-\frac{(\ln M - \mu_0 - \gamma\,t)^2}{2\sigma_0^2}\right)}.
\eeq
The accretion time $t$ may not be a fixed constant that applies to all objects, therefore we can integrate over a function $f(t)$ (which has units of inverse time) that describes the distribution of accretion times. In this case the final observed mass function $f(\ln M) \equiv dn/d\ln M$ is
\beq
\label{master}
 \int_{0}^{T}  \frac{1}{\sqrt{2\pi}\sigma_0}\exp{\left(-\frac{(\ln M - \mu_0 - \gamma\,t')^2}{2\sigma_0^2}\right)}f(t') \, dt'.
\eeq
Here $f(t')$ is a normalized distribution of accretion times $t'$ and $T$ is the maximum possible accretion time.
%The distribution of accretion lifetimes is normalized such that 
%\beq  
%\int_{0}^{T} f(t)\,dt = 1.
%\eeq
The function $f(t')$ is determined by considering the creation rate of black holes in the DCBH scenario. The number density $n$ of black holes grows in a type of chain reaction \citep{yue14,dij14} with the instantaneous creation rate $dn/dt = \lambda(t)\,n$. The simplest case where $\lambda(t) = \lambda$ has a constant value leads to pure exponential growth. If this growth continues from a time $t=0$ when the first DCBH is created until a time $T$ when the creation of all DCBH is terminated, then each black hole that was created at time $t$ has an accretion lifetime $t' = T-t$ in the range $[0,T]$. The normalized distribution of accretion lifetimes $t'$ is then
%If for example the number of black holes is growing exponentially with time $t_B$ since the Big Bang, i.e.,
%\beq
%\log N(t_B) = \nu (t_B-t_0), \mathtt{where}\:  t_0 \leq t_B \leq t_f
%\eeq
%and each black hole accretes from the time it is created until time $t_f$, , then the distribution of accretion lifetimes $t = (t_f-t_0) - t_B$ is 
\beq
\label{lifetimes}
f(t') = \frac{\lambda\,\exp (-\lambda\, t')}{[1-\exp (-\lambda\,T)]}.
\eeq
Using the indefinite integral identity
\begin{eqnarray}
 && \int  \exp[-(ax^2 + bx + c)]dx \nonumber \\
 & = & \frac{1}{2} \sqrt{\frac{\pi}{a}} \exp\, \left(\frac{b^2 - 4ac}{4a}\right) \erf \left(\sqrt{a}\left[x + \frac{b}{2a}\right]\right) ,
\end{eqnarray}
%I_1 =  \int \exp[-(ax^2 + bx + c)]dx \, , \ a > 0 
valid for $a>0$, we evaluate 
%By completing the square and letting $u = \sqrt{a}\left(x+\frac{b}{2a}\right)$ it can be brought to
%
%\beq  
%I_1 =  \frac{1}{2} \sqrt{\frac{\pi}{a}} \exp \left(\frac{b^2 - 4ac}{4a}\right) \erf %\left(\sqrt{a}\left[x + \frac{b}{2a}\right]\right) + \tilde{C}\ ,
%\eeq
%
%where $\tilde{C}$ is an integration constant. 
the integral in Equation (\ref{master}) using Equation (\ref{lifetimes}) and obtain a full expression
%
%\begin{eqnarray}
% f(\ln M) & = & \frac{\alpha\,\exp\,(\alpha\mu_0 +\alpha^2\sigma_0^2 / 2)}{2[1-\exp\,(-\lambda\,T)]}\,M^{-\alpha} \nonumber \\ 
% & \times & [\erf\left(\frac{1}{\sqrt{2}}\left(\alpha\sigma_0 - \frac{\ln M - \mu _0 - \gamma\, T}{\sigma_0}\right) \right) \nonumber \\
% & - &\erf\left(\frac{1}{\sqrt{2}}\left(\alpha\sigma_0 - \frac{\ln M - \mu _0}{\sigma_0}\right)\right) ].
%\end{eqnarray}
\begin{eqnarray}
 f(\ln M) & = & \frac{\alpha\,\exp\,(\alpha\mu_0 +\alpha^2\sigma_0^2 / 2)}{2[1-\exp\,(-\alpha\beta)]}\,M^{-\alpha} \nonumber \\ 
 & \times & \bigg[ \erf\left(\frac{1}{\sqrt{2}}\left(\alpha\sigma_0 - \frac{\ln M - \mu _0 - \beta}{\sigma_0}\right) \right) \nonumber \\
 & - &\erf\left(\frac{1}{\sqrt{2}}\left(\alpha\sigma_0 - \frac{\ln M - \mu _0}{\sigma_0}\right)\right) \bigg].
 \label{tlp}
\end{eqnarray}
Here $\alpha \equiv \lambda/\gamma$, the dimensionless ratio of the growth rate of DCBH formation to the growth rate of the mass of individual DCBH, and $\beta \equiv \gamma\,T$, 
the dimensionless number of DCBH growth times within the DCBH formation era. 

\begin{figure}[h]
	\plotone{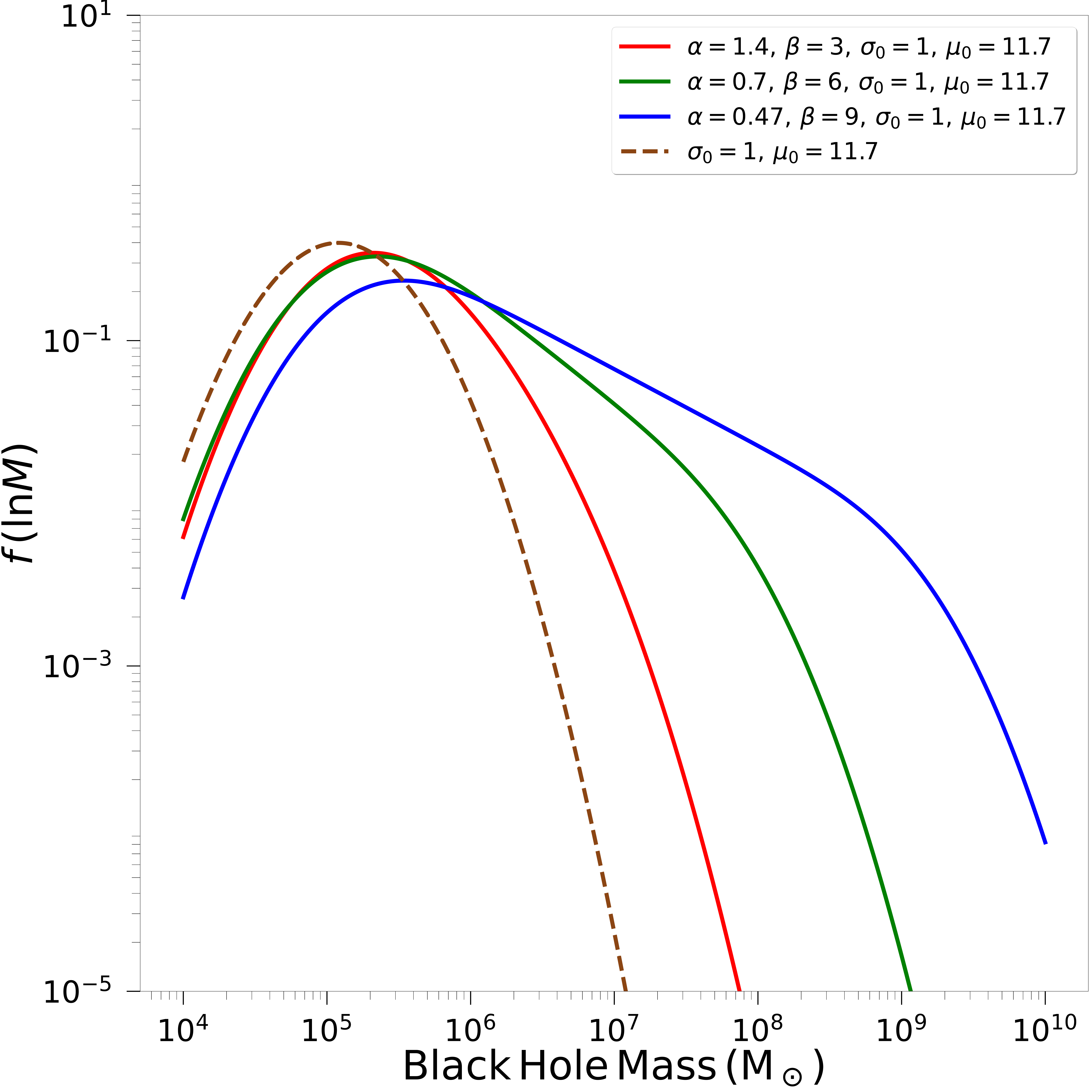}
	\caption{The tapered power law (TPL) distribution and an underlying lognormal distribution. Parameters are chosen as plausible values based on models of the DCBH growth era and also illustrate important features of the distribution. Shown is an underlying lognormal distribution (dashed
		line) with $\mu_0 = 11.7, \sigma_0=1$, and the TPL distributions generated from it assuming either of the following: Eddington-limited growth with $\chi=1$ (red line), super-Eddington growth with $\chi = 2$ (green line), and super-Eddington growth with 
		$\chi = 3$ (blue line). Note that $\mu_0=11.7$ corresponds to a peak mass
	$10^{5.1}\, \Msun$. }
	\label{flnm}
\end{figure}

In the limit $T \rightarrow \infty$, the function becomes
\begin{eqnarray}
 f(\ln M)& = & \frac{\alpha}{2}\exp\,(\alpha\mu_0 +\alpha ^2\sigma_0^2/2)\,M^{-\alpha} \nonumber \\
 &\times& \erfc\left(\frac{1}{\sqrt{2}}\left(\alpha\sigma_0 - \frac{\ln M - \mu _0}{\sigma_0}\right)\right), 
 \label{mlp}
\end{eqnarray}
which is the modified lognormal power law (MLP) function \citep{bas15}. 
Equation (\ref{tlp}) represents a tapered version of the MLP, with the break in the power-law occurring at 
%$\ln M \approx \mu_0 + \beta$, i.e., 
$\log M \approx (\mu_0 + \beta)/2.3$, 
meaning that the peak of the original lognormal is shifted in $\ln M$ by an amount
$\beta = \gamma T$. Henceforth, we refer to Equation (\ref{tlp}) as the tapered power law (TPL) function. 
%The MLP has a high-mass power-law tail of index $\alpha$, while the TPL has a similar form but has a break in the power-law tail that occurs where the peak of the underlying lognormal distribution has shifted by an amount $\beta = \gamma\,T$, hence at a characteristic value $\ln M = \mu_0 + \beta$.

Figure \ref{flnm} shows the TPL function using parameter values obtained from models of
the DCBH growth era. We pick $\mu_0=11.7$ and $\sigma_0=1.0$ based on the model of 
\citet{fer14}. From \citet{dij14} (see Figure \ref{fnum}) we adopt $\lambda=28.0$ Gyr$^{-1}$ 
for the era of rapid DCBH formation using their canonical model. The length 
of the DCBH growth era is $T=0.15$ Gyr \citep{yue14}.
For accretion growth during this period we expect that super-Eddington growth 
can occur \citep{dij14} but with a wide range of possible values. We 
pick a series of values $\chi=[1,2,3]$ covering Eddington-limited growth
and two values of super-Eddington growth. Since $\gamma = \chi\gamma_0 = 20\chi$ Gyr$^{-1}$, this leads to $\alpha = [1.4, 0.7, 0.47]$ and $\beta = [3,6,9]$ 
for our adopted values of $\lambda$ and $T$.
%dimensionless so that $\gamma = \chi/\tedd \simeq 66.7$ Gyr$^{-1}$
%using the canonical value $\tedd = 45$ Myr. This yields $\alpha = \lambda/\gamma
%= 0.43$. Furthermore, using the value $T=150$ Myr for the DCBH growth era \citep{yue14}
%we find $\beta = \gamma T = 10.0$.
Figure \ref{flnm} shows that the super-Eddington growth models allow for the development of 
a mass function that has both a visually evident power-law profile as well as a 
notable break in the power law at high mass. This break is a marker of the end
of the DCBH growth era, since both the creation of new DCBH as well as their
super-Eddington growth ceases after the time interval $T$.

%subsection on typical/expected values of parameters alpha and beta?

\section{The Quasar Luminosity Function}

Once the DCBH growth era has ended at $z \approx 13$, the population of DCBH may continue to 
undergo Eddington-limited accretion, and the luminosity function can be 
estimated using Equation (\ref{Ledd}).
Over time, the mass function $f(\ln M)$ will retain its shape but move to
the right since $\ln M$ at the end of the DCBH era will shift by an amount
$\gamma\Delta t_z$ where $\Delta t_z$ is the time 
interval between the end of the DCBH era ($z \approx 13$) and an observable
redshift $z$. However, the duty cycle $\Duty$ and therefore $\chi = \Duty \fedd$
may be $\ll 1$ in this 
era, rendering mass growth to small fractional levels. A random sampling 
of $\Duty$ and $\fedd$ for individual object growth after $z \approx 13$ shows that
the overall distribution maintains its shape and moves to the right in $\log M$.
We also note that the mass growth of SMBH may be quenched above $\sim 10^{10}\,\Msun$ 
%quenched by star formation in the surrounding disk as well as strong outflows
%and jets 
\citep{ina16,ich17}, in agreement with results of current quasar surveys \citep{ghi10,tra14}.

%We assume that after the period of rapid super-Eddington growth, each DCBH continues
%to grow at an Eddington-limited accretion rate. In this case, we can convert the TPL
%mass function into a luminosity function. The mass function will shift to higher masses
%over time but will do so much more slowly than in the DCBH growth era. 

\begin{figure}[h]
	\plotone{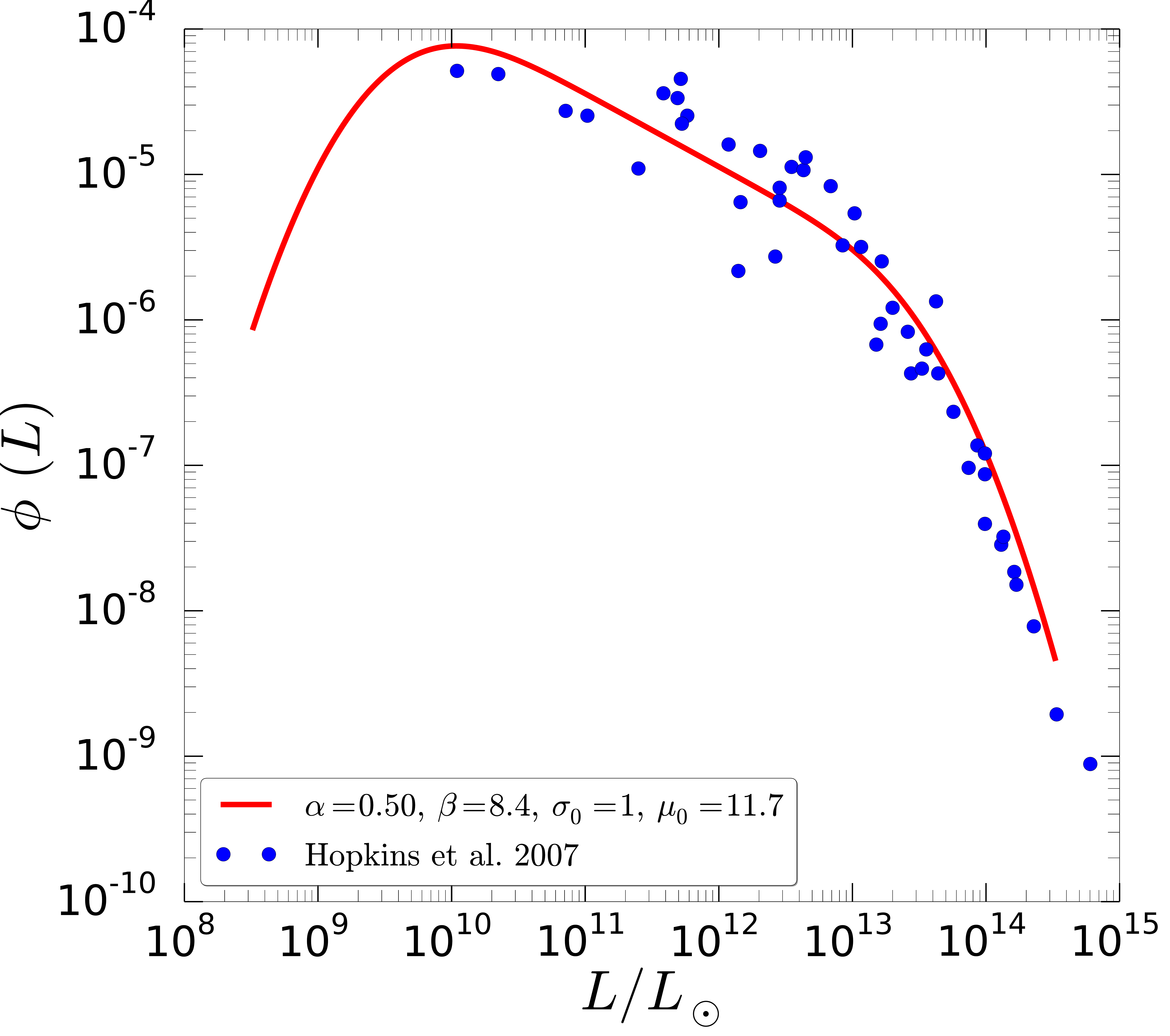}
	\caption{Probability distribution of quasar luminosities. The TPL function is plotted with parameters $\alpha = 0.5$ and $\beta =8.4$ in which $\mu_0$ and $\sigma_0$ are held fixed at 11.7 and 1.0, respectively. Data points are estimates of bolometric 
	luminosity of quasars at $z=3$ taken from \citet{hop07}.}
	\label{flum}
\end{figure}

Assuming that observed quasars are undergoing Eddington-limited accretion, 
we use Equation (\ref{Ledd}) to 
transform the mass function into a luminosity function.
We expect that the mass of the quasars are not growing substantially during this time, 
for reasons discussed above, and we are really most interested in fitting the shape
of the function, which should remain much the same for a variety of
redshifts in the post-DCBH-growth era. Figure \ref{flum} shows the inferred 
quasar luminosity function (QLF) $\phi(L) \propto dn/d \log L$ for a suitable 
pair of values for $\alpha$ and $\beta$, overlaid on $z=3$ quasar bolometric 
luminosities compiled by \citet{hop07}. 
Here we are only interested in fitting the shape of the luminosity function and not the absolute number of sources. The 
normalization can be scaled to fit the observed number at any redshift.
%The time interval between $z=13$ and $z=3$ is about 1.8 Gyr, and we ignore here any evolution in the QLF that may occur during this interval. 
%leaving such 
%higher order fitting considerations to a future paper. 
%Eddington-limited
%accretion is expected to change the absolute values of mass and luminosity
%somewhat, but not affect the shape of their distributions. 
%what is time interval between z=13 and z=3?
%lum fcn with move to right linearly with time

We hold $(\mu_0,\sigma_0)$ fixed at their model-inspired values
$(11.7,1.0)$ since they are most
important in determining the unobserved low end of the luminosity function. 
We effectively fit observations with two parameters $(\alpha,\beta)$. This is 
in contrast to the usual practice of fitting the observed QLF with a double power law
\citep[e.g.,][]{hop07,mas12,sch19} that requires three parameters: the two power-law indices and a joining point. 

% pick up brighter quasars so we can afford to keep we find that 
%$\alpha = 0.50$ and $\beta = 8.4$ provide an excellent fit to the data. The parameters
%$\mu_0$ and $\sigma_0$ are most relevant at low masses and luminosities, in a regime
%that is not easily observable. 

In our model, the values of $\alpha,\beta$ that fit the QLF are not merely mathematical
parameters. Instead, they reveal the history of the putative DCBH growth era. 
For the adopted DCBH number growth rate $\lambda = 28.0$ Gyr$^{-1}$ and a duration $T= 0.15$ Gyr,
%taken from \citet{dij14} and \citet{yue14}, respectively, 
and individual masses growing at a rate $\gamma = \chi \gamma_0$,
the two fitted parameters are related to the super-Eddington factor $\chi$ by
\begin{eqnarray}
	\alpha & = & 1.4 \chi^{-1}, \\ 
	\beta & = & 3 \chi.
	\label{alphabeta}
\end{eqnarray}
We find an excellent fit to the QLF with $[\alpha,\beta] = [0.5,8.4]$. Both the values
of $\alpha$ and $\beta$ imply a super-Eddington factor $\chi = 2.8$, revealing the self-consistency of our model.
In principle the QLF could have been fit with any $\alpha$ and $\beta$ that could individually
imply very different values of $\chi$. In that case our underlying model would be inconsistent,
or at least need to explore values of $\lambda$ and $T$ that were quite different 
than those implied by current models of the DCBH growth era. 

To elaborate on the above point, our model could in principle also 
be applied to SMBH formation from alternate scenarios 
such as Population III remnants \citep{mad01,wha12} or mergers of primordial 
Population III stars \citep{boe18,rei18}. It could apply as long as the black hole
production could be described as growing exponentially at some rate $\lambda$ and
for a finite time $T$, during which the individual masses grew at an Eddington-limited
or super-Eddington rate.

%The values of $\alpha$ and $\beta$ are a self-consistent
%pair, which reinforces the applicability of our model and a convincing link 
%between the QLF and the simulation results of
%a rapid DCBH growth era with super-Eddington accretion \citep[e.g.,]{fer14,yue14,dij14,pac17}.
%Adopting $\lambda = 28.0$ Gyr$^{-1}$, 
%and using $\gamma = 20 \chi$ Gyr$^{-1}$, we find $\alpha = 1.4\chi^{-1}$ so that
%there is an implied value $\chi = 2.8$. 
%This confirms that an epoch of 
%super-Eddington accretion is required. The same value of $\chi=2.8$ yields 
%$\beta = 8.4$ since $\beta = \gamma T = 3 \chi$ and our adopted $T= 0.15$ Gyr. 

\section{Summary}

We have presented an analytic model that captures some essential features of the 
DCBH growth scenario and uses them to derive an analytic mass function and by 
implication a luminosity function. A double power-law function has been commonly
used in the literature to mathematically fit the quasar luminosity function (QLF). 
Here, we instead use a physically-motivated formula based on the 
scenario of the DCBH growth era that has been developed by many researchers. It is
not a double-power law at high mass and luminosity, but rather a tapered power-law.
We believe that the rapid fall off in the QLF at high luminosity is better modeled 
as a tapered part of a power law rather than as a second power law. 
The break point of the power law identifies the end of the era of DCBH creation.

We have fit an observed QLF with a power-law index $\alpha = 0.5$ and the break-point related parameter $\beta=8.4$. These are consistent with  
a period of rapid mass growth of DCBH with super-Eddington factor $\chi=2.8$, 
for a time period $T=150$ Myr
%corresponding to the era between $z \approx 20$ and $z \approx 13$, and a growth
during which the growth rate of the number density $\ndcbh$ was
$\lambda = 28.0$ Gyr$^{-1}$.
In principle, the best fit 
values to QLF data can be used to constrain such theoretical models of 
DCBH growth. 

%Our model captures the history of the DCBH growth through two
%parameters: $\alpha$, the power-law index, and $\beta$, which is related to the break 
%point of the power law. Since both parameters carry physical meaning, the best fit 
%values to QLF data can be used to compare with theoretical models of 
%DCBH growth. 

Our model has two key components. Initially high mass $\sim 10^5\,\Msun$ seeds 
grow rapidly in number during a limited time period in the early universe, since
DCBH formation leads to the emission of LW photons that seed the 
formation of other DCBH. These 
objects also live within gas-rich halos and undergo super-Eddington mass accretion. 
Then at some time both the
formation of DCBH as well as the super-Eddington accretion of the existing DCBH 
comes to a rapid halt due to the photoevaporation of the host halos. What remains
is a tapered power-law (TPL) distribution of masses and therefore also of luminosity
if the observed quasars are undergoing subsequent Eddington-limited accretion. 
Future modeling can relax some of these assumptions, for example the formation 
	of DCBH may continue long enough to outlive the period of rapid (super-Eddington) mass growth, especially if driven by mechanisms other than the LW flux \citep{wis19},
	and the super-Eddington accretion may not apply to all objects \citep{pac17,lat18}. 
	%In these cases, we expect that at least the low-mass end of the mass function
	%will be substantially affected.

\section*{Acknowledgements}
We thank the referee for constructive comments. 
SB was supported by a Discovery Grant from NSERC. 

\bibliography{mybiblio}

\begin{thebibliography}{}
\expandafter\ifx\csname natexlab\endcsname\relax\def\natexlab#1{#1}\fi
\providecommand{\url}[1]{\href{#1}{#1}}

\bibitem[{{Agarwal} {et~al.}(2012){Agarwal}, {Khochfar}, {Johnson}, {Neistein},
  {Dalla Vecchia}, \& {Livio}}]{aga12}
{Agarwal}, B., {Khochfar}, S., {Johnson}, J.~L., {et~al.} 2012, \mnras, 425,
  2854

\bibitem[{{Ba{\~n}ados} {et~al.}(2018){Ba{\~n}ados}, {Venemans},
  {Mazzucchelli}, {Farina}, {Walter}, {Wang}, {Decarli}, {Stern}, {Fan},
  {Davies}, {Hennawi}, {Simcoe}, {Turner}, {Rix}, {Yang}, {Kelson}, {Rudie}, \&
  {Winters}}]{ban18}
{Ba{\~n}ados}, E., {Venemans}, B.~P., {Mazzucchelli}, C., {et~al.} 2018, \nat,
  553, 473

\bibitem[{{Basu} {et~al.}(2015){Basu}, {Gil}, \& {Auddy}}]{bas15}
{Basu}, S., {Gil}, M., \& {Auddy}, S. 2015, \mnras, 449, 2413

\bibitem[{{Begelman} {et~al.}(2008){Begelman}, {Rossi}, \& {Armitage}}]{beg08}
{Begelman}, M.~C., {Rossi}, E.~M., \& {Armitage}, P.~J. 2008, \mnras, 387, 1649

\bibitem[{{Begelman} {et~al.}(2006){Begelman}, {Volonteri}, \& {Rees}}]{beg06}
{Begelman}, M.~C., {Volonteri}, M., \& {Rees}, M.~J. 2006, \mnras, 370, 289

\bibitem[{{Boekholt} {et~al.}(2018){Boekholt}, {Schleicher}, {Fellhauer},
  {Klessen}, {Reinoso}, {Stutz}, \& {Haemmerl{\'e}}}]{boe18}
{Boekholt}, T.~C.~N., {Schleicher}, D.~R.~G., {Fellhauer}, M., {et~al.} 2018,
  \mnras, 476, 366

\bibitem[{{Bromm} \& {Loeb}(2003)}]{bro03}
{Bromm}, V., \& {Loeb}, A. 2003, \apj, 596, 34

\bibitem[{{Chon} {et~al.}(2016){Chon}, {Hirano}, {Hosokawa}, \&
  {Yoshida}}]{cho16}
{Chon}, S., {Hirano}, S., {Hosokawa}, T., \& {Yoshida}, N. 2016, \apj, 832, 134

\bibitem[{{Dijkstra} {et~al.}(2014){Dijkstra}, {Ferrara}, \&
  {Mesinger}}]{dij14}
{Dijkstra}, M., {Ferrara}, A., \& {Mesinger}, A. 2014, \mnras, 442, 2036

\bibitem[{{Fan} {et~al.}(2006){Fan}, {Strauss}, {Richards}, {Hennawi},
  {Becker}, {White}, {Diamond-Stanic}, {Donley}, {Jiang}, {Kim}, {Vestergaard},
  {Young}, {Gunn}, {Lupton}, {Knapp}, {Schneider}, {Brandt}, {Bahcall},
  {Barentine}, {Brinkmann}, {Brewington}, {Fukugita}, {Harvanek}, {Kleinman},
  {Krzesinski}, {Long}, {Neilsen}, {Nitta}, {Snedden}, \& {Voges}}]{fan06}
{Fan}, X., {Strauss}, M.~A., {Richards}, G.~T., {et~al.} 2006, \aj, 131, 1203

\bibitem[{{Ferrara} {et~al.}(2014){Ferrara}, {Salvadori}, {Yue}, \&
  {Schleicher}}]{fer14}
{Ferrara}, A., {Salvadori}, S., {Yue}, B., \& {Schleicher}, D. 2014, \mnras,
  443, 2410

\bibitem[{{Ghisellini} {et~al.}(2010){Ghisellini}, {Della Ceca}, {Volonteri},
  {Ghirland a}, {Tavecchio}, {Foschini}, {Tagliaferri}, {Haardt}, {Pareschi},
  \& {Grindlay}}]{ghi10}
{Ghisellini}, G., {Della Ceca}, R., {Volonteri}, M., {et~al.} 2010, \mnras,
  405, 387

\bibitem[{{Habouzit} {et~al.}(2016){Habouzit}, {Volonteri}, {Latif}, {Dubois},
  \& {Peirani}}]{hab16}
{Habouzit}, M., {Volonteri}, M., {Latif}, M., {Dubois}, Y., \& {Peirani}, S.
  2016, \mnras, 463, 529

\bibitem[{{Hopkins} {et~al.}(2007){Hopkins}, {Richards}, \&
  {Hernquist}}]{hop07}
{Hopkins}, P.~F., {Richards}, G.~T., \& {Hernquist}, L. 2007, \apj, 654, 731

\bibitem[{{Hosokawa} {et~al.}(2011){Hosokawa}, {Omukai}, {Yoshida}, \&
  {Yorke}}]{hos11}
{Hosokawa}, T., {Omukai}, K., {Yoshida}, N., \& {Yorke}, H.~W. 2011, Science,
  334, 1250

\bibitem[{{Ichikawa} \& {Inayoshi}(2017)}]{ich17}
{Ichikawa}, K., \& {Inayoshi}, K. 2017, \apj, 840, L9

\bibitem[{{Inayoshi} \& {Haiman}(2016)}]{ina16}
{Inayoshi}, K., \& {Haiman}, Z. 2016, \apj, 828, 110

\bibitem[{{Latif} {et~al.}(2018){Latif}, {Volonteri}, \& {Wise}}]{lat18}
{Latif}, M.~A., {Volonteri}, M., \& {Wise}, J.~H. 2018, \mnras, 476, 5016

\bibitem[{{Madau} \& {Rees}(2001)}]{mad01}
{Madau}, P., \& {Rees}, M.~J. 2001, \apj, 551, L27

\bibitem[{{Masters} {et~al.}(2012){Masters}, {Capak}, {Salvato}, {Civano},
  {Mobasher}, {Siana}, {Hasinger}, {Impey}, {Nagao}, {Trump}, {Ikeda}, {Elvis},
  \& {Scoville}}]{mas12}
{Masters}, D., {Capak}, P., {Salvato}, M., {et~al.} 2012, \apj, 755, 169

\bibitem[{{Mortlock} {et~al.}(2011){Mortlock}, {Warren}, {Venemans}, {Patel},
  {Hewett}, {McMahon}, {Simpson}, {Theuns}, {Gonz{\'a}les-Solares}, {Adamson},
  {Dye}, {Hambly}, {Hirst}, {Irwin}, {Kuiper}, {Lawrence}, \&
  {R{\"o}ttgering}}]{mor11}
{Mortlock}, D.~J., {Warren}, S.~J., {Venemans}, B.~P., {et~al.} 2011, \nat,
  474, 616

\bibitem[{{Pacucci} {et~al.}(2017){Pacucci}, {Natarajan}, {Volonteri},
  {Cappelluti}, \& {Urry}}]{pac17}
{Pacucci}, F., {Natarajan}, P., {Volonteri}, M., {Cappelluti}, N., \& {Urry},
  C.~M. 2017, \apj, 850, L42

\bibitem[{{Reinoso} {et~al.}(2018){Reinoso}, {Schleicher}, {Fellhauer},
  {Klessen}, \& {Boekholt}}]{rei18}
{Reinoso}, B., {Schleicher}, D.~R.~G., {Fellhauer}, M., {Klessen}, R.~S., \&
  {Boekholt}, T.~C.~N. 2018, \aap, 614, A14

\bibitem[{{Sakurai} {et~al.}(2016){Sakurai}, {Vorobyov}, {Hosokawa}, {Yoshida},
  {Omukai}, \& {Yorke}}]{sak16}
{Sakurai}, Y., {Vorobyov}, E.~I., {Hosokawa}, T., {et~al.} 2016, \mnras, 459,
  1137

\bibitem[{{Schindler} {et~al.}(2019){Schindler}, {Fan}, {McGreer}, {Yang},
  {Wang}, {Green}, {Fynbo}, {Krogager}, {Green}, {Huang}, {Kadowaki}, {Patej},
  {Wu}, \& {Yue}}]{sch19}
{Schindler}, J.-T., {Fan}, X., {McGreer}, I.~D., {et~al.} 2019, \apj, 871, 258

\bibitem[{{Shang} {et~al.}(2010){Shang}, {Bryan}, \& {Haiman}}]{sha10}
{Shang}, C., {Bryan}, G.~L., \& {Haiman}, Z. 2010, \mnras, 402, 1249

\bibitem[{{Trakhtenbrot}(2014)}]{tra14}
{Trakhtenbrot}, B. 2014, \apj, 789, L9

\bibitem[{{Vorobyov} {et~al.}(2013){Vorobyov}, {DeSouza}, \& {Basu}}]{vor13}
{Vorobyov}, E.~I., {DeSouza}, A.~L., \& {Basu}, S. 2013, \apj, 768, 131

\bibitem[{{Whalen} \& {Fryer}(2012)}]{wha12}
{Whalen}, D.~J., \& {Fryer}, C.~L. 2012, \apj, 756, L19

\bibitem[{{Wise} {et~al.}(2019){Wise}, {Regan}, {O'Shea}, {Norman}, {Downes},
  \& {Xu}}]{wis19}
{Wise}, J.~H., {Regan}, J.~A., {O'Shea}, B.~W., {et~al.} 2019, \nat, 566, 85

\bibitem[{{Woods} {et~al.}(2018){Woods}, {Agarwal}, {Bromm}, {Bunker}, {Chen},
  {Chon}, {Ferrara}, {Glover}, {Haemmerle}, {Haiman}, {Hartwig}, {Heger},
  {Hirano}, {Hosokawa}, {Inayoshi}, {Klessen}, {Kobayashi}, {Koliopanos},
  {Latif}, {Li}, {Mayer}, {Mezcua}, {Natarajan}, {Pacucci}, {Rees}, {Regan},
  {Sakurai}, {Salvadori}, {Schneider}, {Surace}, {Tanaka}, {Whalen}, \&
  {Yoshida}}]{woo18}
{Woods}, T.~E., {Agarwal}, B., {Bromm}, V., {et~al.} 2018, arXiv e-prints,
  arXiv:1810.12310

\bibitem[{{Wu} {et~al.}(2015){Wu}, {Wang}, {Fan}, {Yi}, {Zuo}, {Bian}, {Jiang},
  {McGreer}, {Wang}, {Yang}, {Yang}, {Thompson}, \& {Beletsky}}]{wu15}
{Wu}, X.-B., {Wang}, F., {Fan}, X., {et~al.} 2015, \nat, 518, 512

\bibitem[{{Yue} {et~al.}(2014){Yue}, {Ferrara}, {Salvaterra}, {Xu}, \&
  {Chen}}]{yue14}
{Yue}, B., {Ferrara}, A., {Salvaterra}, R., {Xu}, Y., \& {Chen}, X. 2014,
  \mnras, 440, 1263

\end{thebibliography}

\end{document}